\documentclass[%
 aps,prl,reprint,
 nofootinbib,
 amsmath,amssymb
]{revtex4-2}

\usepackage{graphicx}
\usepackage{dcolumn}
\usepackage{bm}
\usepackage[colorlinks, citecolor={blue!70!black}, urlcolor={blue!70!black}, linkcolor={red!70!black},hyperindex,breaklinks]{hyperref}
\usepackage{svg}
\usepackage{subfigure}
\usepackage{booktabs}
\usepackage[mathlines]{lineno}

\newcommand{\vdg}{\vphantom{\dagger}}

\begin{document}

\preprint{APS/123-QED}


\title{Apparent Planckian scattering from local polaron formation}

\author{Brian Yong-Ho Lee}
\affiliation{Department of Physics and Astronomy, University of Rochester,  Rochester, New York 14627, USA}

\author{Chaitanya Murthy}
\affiliation{Department of Physics and Astronomy, University of Rochester,  Rochester, New York 14627, USA}

\date{\today}

\begin{abstract}
We propose a simple mechanism for \textit{apparent Planckian scattering} based on local polaron formation, in which $\Gamma_\text{tr} = \Gamma_0 + \alpha k_BT / \hbar$ with $\alpha \sim O(1)$ emerges from quasielastic scattering without fine tuning. 
We provide evidence for our proposal in Monte Carlo simulations of the Holstein model with disordered electron-phonon coupling in the adiabatic limit.
Our mechanism generates a finite interval of couplings in which the slope $\alpha$ is approximately constant, coinciding with the onset of local polaron formation.
In this regime, Matthiessen's rule is dramatically violated---or obeyed, depending on one's point of view---in that changes to the couplings, which in perturbation theory would alter the slope $\alpha$, instead change the intercept $\Gamma_0$.
We conjecture that a version of our mechanism applies to any system with a dominant disordered interaction that can drive polaron formation.
This potentially includes regimes of several recently introduced strange metal models based on spatially random interactions.
\end{abstract}

\maketitle

A variety of materials, from conventional metals to strange metals in the cuprates and other strongly correlated systems, exhibit electrical resistivities that correspond to $T$-linear transport scattering rates $\Gamma_{\text{tr}} \approx \alpha k_B T/ \hbar$ with $\alpha \sim O(1)$~\cite{bruin_similarity_2013, legros_universal_2019, licciardello_electrical_2019, cao_strange_2020, grissonnanche_linear-temperature_2021, jaoui_quantum_2022, lee_linear--temperature_2023, zhakina_investigation_2023}. 
Such scattering rates $\Gamma \approx k_BT/\hbar$ have been dubbed ``Planckian'', especially when they emerge from inelastic scattering processes and persist over a range of parameters~\cite{zaanen_why_2004, hartnoll_colloquium_2022}. 
It is possible that Planckian scattering is intimately related to non-quasiparticle transport, quantum criticality, fundamental bounds on dissipation, and understanding high-$T_c$ superconductivity \cite{phillips_stranger_2022, hartnoll_colloquium_2022}. 
Consequently, a large literature has amassed on observing and explaining the phenomenology of Planckian transport in strange metals 
\cite{zaanen_why_2004, cooper_anomalous_2009,
bruin_similarity_2013, legros_universal_2019, licciardello_electrical_2019, cao_strange_2020, grissonnanche_linear-temperature_2021, jaoui_quantum_2022,
lee_linear--temperature_2023,
zhakina_investigation_2023, 
hartnoll_theory_2015, 
mousatov_theory_2020,
else_strange_2021,
hardy_enhanced_2025,
patel_strange_2025,
aldape_solvable_2022,
patel_universal_2023,
bashan_tunable_2024,
tulipman_solvable_2024,
das_sarma_strange_2022,
hartnoll_colloquium_2022, phillips_stranger_2022, 
ciuchi_strange_2023,
aydin_quantum_2024,
lee_model_2024,
delacretaz_bound_2025,
qi_planckian_2026,
fratini_minimal_2026, 
bashan_extended_2026,
chowdhury_information_2026}.

In conventional metals, $T$-linear resistivity occurs due to electron-phonon scattering in the regime $k_B T \gtrsim \hbar \omega_0/3$, where $\omega_0$ is a characteristic phonon frequency, and $\alpha \approx 2\pi \lambda$, where $\lambda$ is a dimensionless electron-phonon coupling constant. 
At high temperatures scattering by phonons is quasi-elastic, and hence Planckian considerations are \textit{a priori} unimportant. 
Previous work~\cite{murthy_stability_2023} argued that, instead, local polaron%
\footnote{We use ``polaron'' to mean both ``polaron'' and ``bipolaron'', as appropriate.}
formation precludes metallic transport with large values of $\lambda$~\cite{esterlis_breakdown_2018, esterlis_pseudogap_2019, fratini_displaced_2021, chakraverty_possibility_1979, freericks_holstein_1993, benedetti_holstein_1998, alexandrov_breakdown_2001, meyer_gap_2002, capone_polaron_2003, bauer_quantitative_2011, chubukov_eliashberg_2020, yuzbashyan_migdal-eliashberg_2022, chubukov_breakdown_2026, prokofev_limits_2026, fratini_transient_2016}, leading to an effective ``stability bound'' on $\alpha$ that has the same functional form as a Planckian bound but a very different physical origin, and conjectured that similar bounds may apply more generally. Behavior consistent with the idea of a stability bound has been observed in recent theoretical models of strange metals~\cite{patel_strange_2025, hardy_enhanced_2025}. 

In this work, we propose a mechanism for \emph{apparent Planckian scattering} based on local polaron formation, by which the slope of the $T$-linear scattering rate $\Gamma_{\text{tr}} = \Gamma_0 + \alpha k_B T / \hbar$ saturates to $\alpha \sim O(1)$ over a range of couplings without fine-tuning. We conjecture that our mechanism (or a suitably generalized version of it) applies to any metallic system of fermions coupled to bosons, provided the coupling is spatially disordered, dominates the dynamics, and is sufficient to drive local polaron formation and self-trapping. 
We support our conjecture by explicitly verifying the mechanism in numerically exact simulations of a version of the paradigmatic Holstein model with quenched disordered electron-phonon coupling in the limit of zero phonon frequency, $\omega_0 = 0$. 

\begin{figure*}
    \centering
    \includegraphics[width=1.0\linewidth]{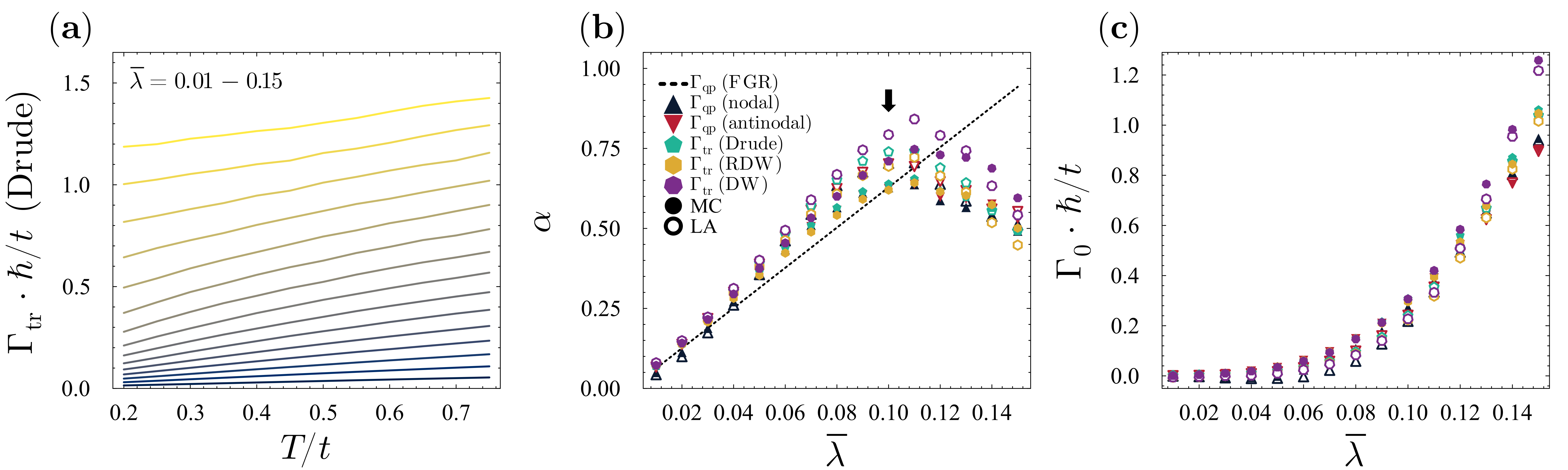}
    \caption{Scattering rates $\Gamma = \alpha T + \Gamma_0$ for the 2D Holstein model with static phonons and disordered $\gamma_i \sim \text{Uniform}(-\gamma_{\text{max}}, \gamma_{\text{max}})$, as a function of average dimensionless bare coupling $\overline{\lambda} = \frac{1}{3} \gamma_{\text{max}}^2 \, \mathcal{N}(0) / K$. 
    \textbf{(a)} Drude scattering rate as a function of temperature $T$, where $\overline{\lambda}$ increases from bottom to top curves.
    \textbf{(b)} $T$-linear slopes $\alpha$ of scattering rates (see text for definitions; FGR = Fermi's golden rule, DW = Drude weight, RDW = renormalized Drude weight) as a function of $\overline{\lambda}$.
    \textbf{(c)} Extrapolated $y$-intercepts $\Gamma_0$ as a function of $\overline{\lambda}$. 
    Solid/open markers represent results computed via exact Monte Carlo (MC) simulations or in the local approximation (LA; see End Matter A), respectively. 
    The solid arrow in panel (b) marks the onset of local polaron formation, diagnosed by a bimodal distribution of phonon displacements on the strongest coupled site (see End Matter A).
    The results presented are for $L = 20$, $t = 1$, $t' = -0.3t$, and $\nu = 1.4$.}
    \label{fig:alpha}
\end{figure*}

The results are summarized in Fig.~\ref{fig:alpha}.
They reveal a finite range of couplings over which the $T$-linear slope $\alpha \sim O(1)$ is approximately independent of the coupling, coinciding with the onset of polaron formation. 
In this regime, Matthiessen's rule is dramatically violated: further changes to the coupling, which in perturbation theory would alter the slope $\alpha$, instead change the intercept $\Gamma_0$.
Nevertheless, Matthiessen's rule would appear to hold \emph{phenomenologically}, since $\Gamma_{\text{tr}} = \Gamma_0 + \alpha k_B T / \hbar$ with $\Gamma_0$ varying and $\alpha$ constant.

We emphasize that our mechanism is not meant to be an explanation for strange metallicity.
However, we speculate that similar ideas may be relevant in a wide class of theoretical models of strange metals based on spatially random interactions \cite{aldape_solvable_2022, patel_universal_2023, patel_strange_2025, bashan_tunable_2024, tulipman_solvable_2024}. 

\vspace{2mm}
\textbf{Caricature of the mechanism.}~%
A caricature of our mechanism, meant to provide intuition, is as follows. 
Consider a lattice system of fermions and adiabatic bosons with fermion-boson interactions on each site parameterized by dimensionless couplings $\lambda_i$.
Assume that the interaction drives the formation and self-trapping of polarons beyond a critical weak-to-intermediate $\lambda_c$.
Now suppose that the $\lambda_i$'s are randomly drawn from a distribution, whose width we will vary.
If the largest coupling $\lambda_{\text{max}} < \lambda_c$, then no polarons form and most sites will be in the weak coupling regime, so the scattering rate is determined by the average coupling, $\Gamma \sim 2\pi \overline{\lambda} T$.
If $\lambda_{\text{max}} > \lambda_c$, however, polarons form and self-trap on sites with $\lambda_i > \lambda_c$ while the remaining sites with $\lambda_i < \lambda_c$ can still be treated in weak coupling. 
The trapped polarons scatter the itinerant carriers like static disorder, but also tend to block them from the strongly-coupled sites.
The scattering rate is therefore $\Gamma \sim \Gamma_0 + 2\pi \overline{\lambda}_c T$, where $\Gamma_0$ is from scattering off the trapped polarons, and $\overline{\lambda}_c$ is the \emph{average coupling on the weak sites} (with $\lambda_i < \lambda_c$).
Crucially, as the distribution of couplings broadens, the $T$-linear contribution to the scattering rate goes from dependent to independent of $\overline{\lambda}$ as $\lambda_{\text{max}}$ crosses $\lambda_c$.

\section*{Model, methods, and observables}
\label{sec:model}

\textbf{Model.}~%
We use natural units $\hbar = k_B = e = 1$. We work with the Holstein model, which describes a band of electrons interacting with a dispersionless optical phonon branch:
\begin{equation}
\begin{aligned}
    H = & - \sum_{ij\sigma} t_{ij}^{\vdg} c_{i\sigma}^\dagger c_{j\sigma}^{\vdg}
    + \sum_i \biggl( \frac{p_i^2}{2M} + \frac{1}{2} K x_i^2 \biggr) \\
    & +\sum_i \gamma_i x_i (n_i - \nu) ,
    \label{H}
\end{aligned}
\end{equation}
where $c_{i\sigma}^\dagger$ creates an electron on site $i$ with spin $\sigma$, $t_{ij}$ is the hopping matrix element between sites $i$ and $j$, $x_i$ and $p_i$ are the phonon displacement and momentum on site $i$, $\omega_0 = \sqrt{K/M}$ is the phonon frequency, $n_i = \sum_\sigma c^\dagger_{i\sigma} c^{\vdg}_{i\sigma}$ is the local electron density, $\nu$ is the (spinful) filling factor, and $\gamma_i$ is the electron-phonon coupling constant on site $i$.
In the standard Holstein model, $\gamma_i = \gamma$ is uniform.
In our version of the model, the $\gamma_i$'s are instead i.i.d.~random variables with a probability distribution $Q(\gamma)$.%
\footnote{When $\gamma_i$ is spatially varying, coupling $x_i$ to the electron density $n_i$ is \emph{not equivalent} to coupling it to the deviation $n_i - \nu$ (unlike in the uniform $\gamma_i = \gamma$ harmonic case, where the difference is just a uniform shift of the phonon coordinates and chemical potential).}
Only recently has \eqref{H} been studied with spatially-varying coupling, but in quite different contexts of metallic nanostructures~\cite{bose_electron_2026} and stabilizing a supersolid phase~\cite{meng_supersolid_2024}.

We solve the model in the adiabatic limit $M \to \infty$ (implying $\omega_0 \to 0$) with $K$ finite, for two reasons. First, in this limit \eqref{H} can be mapped onto an ensemble of non-interacting Hamiltonians with quenched and annealed disorder that can be solved using classical Monte Carlo, circumventing the need to perform analytical continuation to compute dynamical observables like the optical conductivity~\cite{ 
esterlis_breakdown_2018, esterlis_pseudogap_2019, murthy_stability_2023}. Second, the regime of $T$-linear resistivity in the Holstein model is $T \gtrsim \omega_0$, so taking $\omega_0 \to 0$ allows us to focus on the mechanism for apparent Planckian scattering in the relevant temperature range for the model. 
Previous studies have verified that results for various thermodynamic observables in the temperature range of interest are unchanged if calculations are carried out with finite $\omega_0$~\cite{esterlis_breakdown_2018, esterlis_pseudogap_2019}.

\vspace{2mm}
\textbf{Methods.}~%
In the adiabatic limit, disorder-averaged thermal expectation values of any electronic observable $\mathcal{O}_\text{e}$ can be computed as
\begin{equation}
    \overline{\langle \mathcal{O}_\text{e} \rangle} =
    \int d\vec{\gamma} \ Q(\vec{\gamma}) 
    \int d\vec{x} \, P(\vec{x}; \vec{\gamma}) \, \mathcal{O}_\text{e}(\vec{x}; \vec{\gamma}) ,
    \label{eq:Oe_average}
\end{equation}
where $Q(\vec{\gamma}) = \prod_i Q(\gamma_i)$,
\begin{equation}
    P(\vec{x}; \vec{\gamma}) \propto 
    e^{-(\beta K/2) \sum_i(x_i - \gamma_i\nu/K)^2 + \log Z_\text{e}(\vec{x}; \vec{\gamma})} ,
\end{equation}
\begin{equation}
    \mathcal{O}_\text{e}(\vec{x}; \vec{\gamma}) = 
    \text{Tr}_\text{e}\Bigl[e^{-\beta H_\text{e}(\vec{x}; \vec{\gamma}) - \beta \mu N} \, \mathcal{O}_\text{e}\Bigr] / Z_\text{e}(\vec{x}; \vec{\gamma}) ,
\end{equation}
\begin{equation}
    H_\text{e}(\vec{x}; \vec{\gamma}) = -\sum_{ij\sigma} t_{ij} c_{i\sigma}^\dagger c_{j\sigma}^{\vdg} + \sum_i \gamma_i x_i n_i ,
\end{equation}
$Z_\text{e}(\vec{x}; \vec{\gamma}) = \text{Tr}_\text{e}[e^{-\beta H_\text{e}(\vec{x}; \vec{\gamma}) + \beta \mu N}]$, and $N = \sum_{i} n_{i}$, with $\mu$ obtained by solving $\overline{\langle N \rangle} / V = \nu$, where $V$ is the total number of lattice sites. We sample phonon configurations $\vec{x}$ from $P(\vec{x};\vec{\gamma})$ using classical Monte Carlo, compute $\mathcal{O}_\text{e}(\vec{x};\vec{\gamma})$ using exact diagonalization, and perform a quenched disorder average with respect to $Q(\vec{\gamma})$ to estimate $\overline{\langle \mathcal{O}_\text{e} \rangle}$. Further details on the Monte Carlo (MC) algorithm are provided in Ref.~\cite{murthy_stability_2023} and Appendix~\ref{app:MALA+ED}. 

We also solve the model using a local approximation (LA) in which the full joint distribution $P(\vec{x}; \vec{\gamma})$ of the phonon displacements is approximated as a product of suitably chosen local distributions $\prod_i P_i(x_i; \vec{\gamma})$, simplifying \eqref{H} into a quenched disorder problem. 
Details on the local approximation are provided in End Matter A.

\vspace{2mm}
\textbf{Key parameters.}~
The important dimensionless parameters in the problem are the local \textit{bare couplings}
\begin{equation}
    \lambda_i = \gamma_i^2 \mathcal{N}(0)/K ,
\end{equation}
where $\mathcal{N}(0)$ is the bare electronic density of states per spin at the chemical potential, and their disorder average
\begin{equation}
    \overline{\lambda} = \overline{\gamma^2} \mathcal{N}(0)/K ,
\end{equation}
where $\overline{\gamma^2} = \int d\gamma \, Q(\gamma) \, \gamma^2$. Hereafter, ``coupling" will always refer to the bare couplings $\lambda_i$ or $\overline{\lambda}$ unless otherwise stated.
In the clean Holstein model, there is a non-universal but reasonably well defined \textit{critical coupling} $\lambda_\star^\text{clean}$ around which polarons form and self-trap, precipitating a metal-to-insulator crossover \cite{esterlis_breakdown_2018, murthy_stability_2023}.
On the two-dimensional square lattice $\lambda_\star^\text{clean}  \approx 0.3-0.4$.%
\footnote{Strictly speaking there is no single value of the coupling below which no polaron forms and above which a polaron forms, which is why we will speak in approximate terms.} 
In contrast, increasing the coupling $\overline{\lambda}$ in the disordered system sequentially triggers polaron formation at sites with $\lambda_i \gtrsim \lambda_\star^\text{clean}$. This gradual accumulation of self-trapped polarons allows for a sustained metallic phase of itinerant carriers coexisting with local polarons.

\vspace{2mm}
\textbf{Observables.}~%
The primary quantities of interest in this work are the transport and quasiparticle scattering rates, $\Gamma_{\text{tr}}$ and $\Gamma_{\text{qp}}$.
We extract these rates from the longitudinal optical conductivity, $\sigma(\omega)$, and the electron spectral function, $A(\bm{k},\omega)$, respectively.

We compute the longitudinal optical conductivity using the Kubo formula
\begin{equation}
\begin{aligned}
    \sigma(\omega; \vec{x}, \vec{\gamma}) = 
    \frac{2\pi}{\omega V}\sum_{nm} \, 
    &\lvert J_{nm} \lvert^2  \, \bigl[ n_F(\varepsilon_n) - n_F(\varepsilon_{m}) \bigr] \\
    &\times \delta(\varepsilon_n - \varepsilon_{m} + \omega) ,
\end{aligned}
\end{equation}
where $J_{nm} = \langle n \lvert J \lvert m \rangle$ are matrix elements of the current operator in the spinless single-particle eigenstate basis of $H_\text{e}(\vec{x}; \vec{\gamma})$, with energies $\varepsilon_n$, and $n_F(\varepsilon)$ is the Fermi function. The factor of $2$ accounts for spin degeneracy. 
We compute the electron spectral function as 
\begin{equation}
    A(\bm{k}, \omega; \vec{x}, \vec{\gamma}) = 
    \sum_{n} \, \lvert \langle n \lvert \bm{k} \rangle \lvert^2 \, 
    \delta(\omega - \varepsilon_n) ,
\end{equation}
where $\lvert \bm{k} \rangle$ is a single-particle quasimomentum eigenstate. 
The observables are averaged over phonon configurations $\vec{x}$ (annealed) and couplings $\vec{\gamma}$ (quenched) as in Eq.~\eqref{eq:Oe_average} to obtain $\sigma(\omega)$ and $A(\bm{k}, \omega)$. We omit spatial indices on $\sigma(\omega)$ because it is isotropic after the disorder-averaging.

We use three methods from the existing experimental and theoretical literature on Planckian scattering to extract scattering rates from $\sigma(\omega)$ and $A(\bm{k}, \omega)$. Each method produces approximately $T$-linear scattering rates $\Gamma$ in the parameter regime studied (Appendix~\ref{app:gamma_T}). Performing a linear-in-$T$ fit to each extracted $\Gamma(T, \overline{\lambda})$ yields the $T$-linear slope $\alpha(\overline{\lambda})$ and $y$-intercept $\Gamma_0(\overline{\lambda})$. 

\vspace{1mm}
(i) \textit{Drude fit.}
We fit $\sigma(\omega)$ to the Drude form $\sigma_\text{D}(\omega) = \pi D \, \Gamma_\text{tr} / (\Gamma_\text{tr}^2 + \omega^2)$ and extract $\Gamma_\text{tr}$ (and $D$) as fitting parameters.
The resulting rate is reported as $\Gamma_{\text{tr}}^{\text{Drude}}$.

\vspace{1mm}
(ii) \textit{Drude weight.}
%
%
We compute the Drude weight $D_0$ of the bare tight-binding Hamiltonian by the optical sum rule, $D_0 = \langle \tau \rangle_0/V$, where $\langle \tau \rangle_0$ is the expectation value of the diamagnetic stress tensor in the bare equilibrium state. Then, we calculate $\Gamma_{\text{tr}}^{\text{DW}} = D_0 / \sigma(0)$. We also compute the \textit{renormalized Drude weight} $D = \langle \tau \rangle / V$, where $\langle \tau \rangle$ is the expectation value in the exact equilibrium state, and calculate $\Gamma^\text{RDW}_\text{tr} = D/\sigma(0)$.

\vspace{1mm}
(iii) \textit{Nodal/antinodal energy distribution curves.} We compute the spectral function $A(\bm{k}, \omega)$ evaluated near the ``nodal'' ($k_x = k_y$) or ``antinodal'' ($k_y = 0$) Fermi surface $\bm{k}$-points of the bare tight-binding Hamiltonian. Then, we fit the resulting energy distribution curves to the usual form of the on-shell spectral function near the (remains of the) Fermi surface, $A_a(\omega) = (Z_a/2\pi) \, \Gamma_{\text{qp}}^a / [(\Gamma_{\text{qp}}^a/2)^2 + \omega^2]$, where $a = \text{nodal}/\text{antinodal}$, and extract $\Gamma_{\text{qp}}^a$ (and $Z_a$) as fitting parameters.

\vspace{1mm}
Further details about the spectral function, optical conductivity, and the scattering rate extraction methods are provided in Appendix~\ref{app:spec_opcond}.

\begin{figure*}
    \centering
    \includegraphics[width=1.0\linewidth]{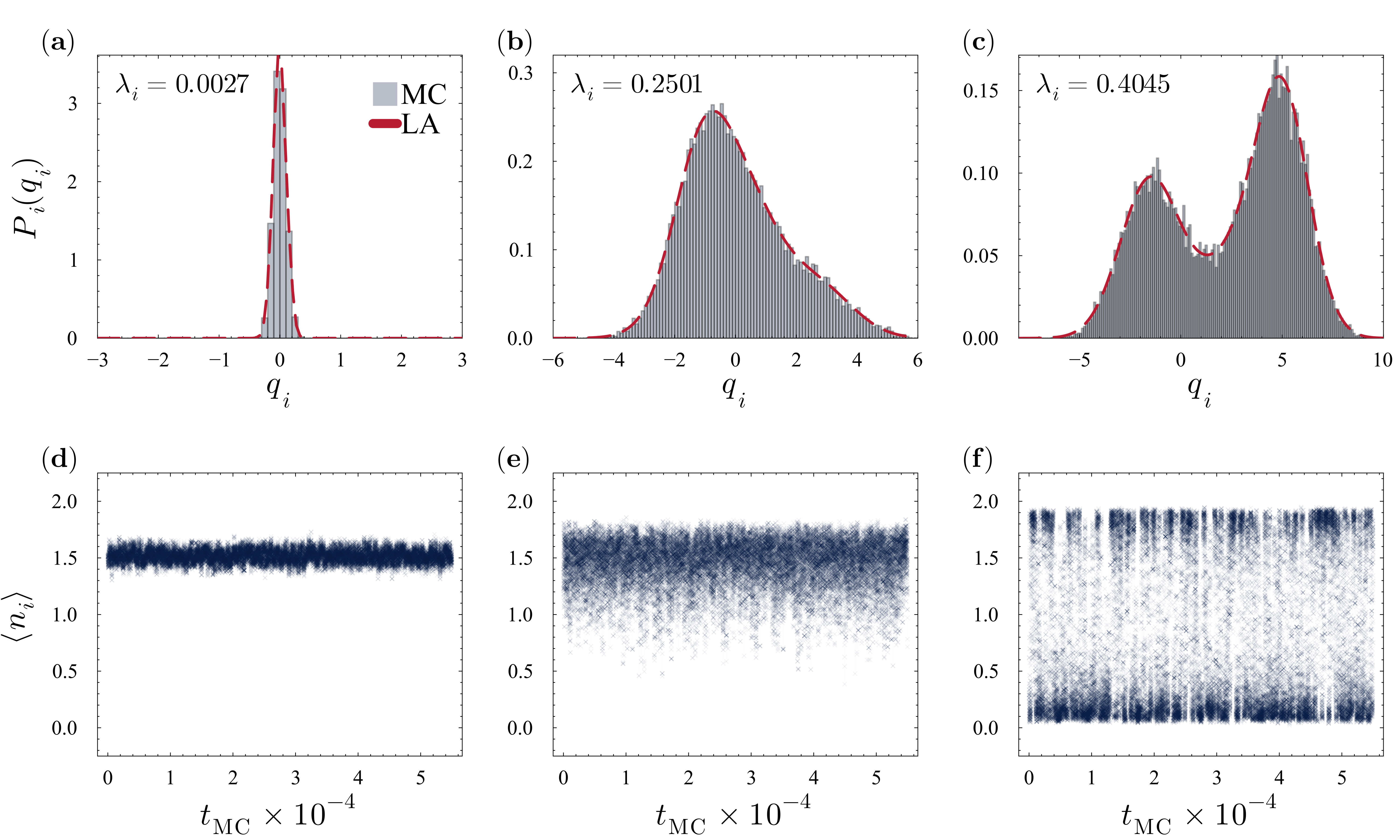}
    \caption{Distributions of effective site potentials $q_i = \gamma_i x_i$ (a--c) and Monte Carlo traces of the local electron density $\langle n_i \rangle$ (d--f) on representative sites with weak (a, d), intermediate (b, e), and strong (c, f) coupling $\lambda_i$ at temperature $T = 0.4t$.
    Other model parameters are the same as in Fig.~\ref{fig:alpha}. 
    Panels (a--c) show the distributions measured in Monte Carlo (MC) as well as fits to the MC using the local approximation (LA).}
    \label{fig:p(q)}
\end{figure*}

\section*{Results}
\label{sec:results}

Representative results for the $T$-linear scattering rates computed from MC and the LA are shown in Fig.~\ref{fig:alpha} for the adiabatic Holstein model with uniformly distributed electron-phonon couplings $\gamma_i \sim \text{Uniform}(-\gamma_{\text{max}}, \gamma_{\text{max}})$, where $\lambda_{\text{max}} = \gamma_{\text{max}}^2 \, \mathcal{N}(0)/K$ and $\overline{\lambda} = \lambda_{\text{max}}/3$. 
We have verified that the key features of the results do not depend sensitively on the particular choice of parameters or model details, such as the filling, band structure, or the precise form of the coupling distribution $Q(\gamma)$ (see End Matter B). For all MC simulations, we use $5000$ burn steps and $55000$ sampling steps. When appropriate, we perform a quenched disorder average over $10$ realizations of the couplings.

\vspace{2mm}
\textbf{Saturation of $T$-linear slopes of scattering rates.}
In the weak coupling regime, the slopes agree with Fermi's golden rule (FGR), $\alpha = 2\pi \overline{\gamma^2} \mathcal{N}(0)/K = 2\pi \overline{\lambda}$. 
We find a critical coupling $\overline{\lambda}_\star \approx 0.1$ around which all the slopes begin to plateau (keep in mind that this corresponds to~$\lambda_{\text{max}}^{\star} = 3\overline{\lambda}_\star \approx 0.3$). 
At larger values of $\overline{\lambda}$, before an eventual metal--insulator crossover, the system enters an apparent Planckian scattering regime in which the slope of the scattering rate takes on a relatively coupling-independent value $\alpha \sim \mathcal{O}(1)$.
%
%
The $y$-intercept of the scattering rate, $\Gamma_0$, remains close to zero until the Planckian scattering regime in which it rapidly grows as a function of $\overline{\lambda}$.

\vspace{2mm}
\textbf{Local polaron formation.}~%
We find that polarons begin to form around the same coupling $\overline{\lambda}_\star \approx 0.1$ ($\lambda^{\text{max}}_{\star} \approx 0.3$) at which $\alpha$ starts to saturate. Polaron formation on a site $i$ is indicated by the emergence of a bimodal distribution for the effective site potential $q_i = \gamma_i x_i$ and thus electron densities tending to the values $\langle n_i \rangle \in \{0, 2\}$~\cite{millis_fermi-liquid--polaron_1996, ciuchi_charge-ordered_1999}. 
Representative distributions $P_i(q_i)$ on sites with weak, $\lambda_i \ll \lambda^{\text{max}}_{\star}$, intermediate, $\lambda_i \lesssim \lambda^{\text{max}}_{\star}$, and strong, $\lambda_i > \lambda^{\text{max}}_{\star}$, coupling for a single coupling realization are shown in Fig.~\ref{fig:p(q)}a--c. 
The weak coupling distribution is unimodal and Gaussian while the strong coupling distribution is bimodal. 
The intermediate coupling distribution remains unimodal but displays significantly non-Gaussian features. 
The coupling $\bar{\lambda}$ at which the first bimodal distribution $P_i(q_i)$ appears is indicated by the bold arrow in Fig.~\ref{fig:alpha}b (see End Matter A), showing that the
formation of polarons is concomitant with the saturation of $\alpha$. Monte Carlo traces of the local electron density on the same sites are shown in Fig.~\ref{fig:p(q)}d--f. 
For weak coupling, the electron density fluctuates around the homogeneous value $\langle n_i \rangle \approx \nu$. 
As the coupling increases, the density tends to fluctuate between $\langle n_i \rangle \approx 0$ and $\langle n_i \rangle \approx 2$.%
\footnote{We remark that this bimodality and statistical switching behavior on strongly coupling sites is reminiscent of local two-level systems, the \textit{dynamics} of which play a key role in the strange metal mechanisms of Refs.~\cite{bashan_tunable_2024, tulipman_solvable_2024}.}

\vspace{2mm}
\textbf{Validity of the local approximation.}~%
Fits to the effective site distributions, which determine the LA, are shown as dashed lines in the top panels of Fig.~\ref{fig:p(q)}. 
The slopes and $y$-intercepts computed in the LA agree qualitatively with the exact results as shown in Fig.~\ref{fig:alpha}.

\section*{Discussion}
\label{sec:discussion}

\vspace{2mm}
\textbf{Synthesis of the results.}~%
Because the LA successfully reproduces the apparent Planckian scattering regime, spatial correlations between phonon configurations are irrelevant to this phenomenon. This indicates that apparent Planckian scattering must be explainable entirely through local physics. Examining individual sites, we observe a Gaussian distribution of phonon coordinates for weak coupling, which transitions into a bimodal distribution characteristic of polaron formation above a critical coupling. Crucially, the $T$-linear slope $\alpha$ only begins to saturate when $\overline{\lambda} \approx 0.1$ where the bimodal, polaronic features also first emerge. We conclude that the saturation of $\alpha$ is a direct consequence of local polaron formation.

\vspace{2mm}
\textbf{The mechanism.}~%
Our results paint the following physical picture. At weak coupling, all carriers are itinerant and scatter according to Fermi's golden rule, $\Gamma(T) = 2\pi \overline{\gamma^2} \mathcal{N}(0)T/K = 2\pi \overline{\lambda} T$. Near the critical coupling, sites with $\lambda_i \gtrsim \lambda_\star$ begin to take on a polaronic character, indicated by the bimodal features in the local phonon distribution. By nature, these particle and hole polaronic sites are less accessible to the remaining itinerant carriers due to Pauli blocking or a high energy barrier. Consequently, the phonon configuration on a polaronic site partially decouples from the itinerant carriers and contributes less to the $T$-linear part of the scattering rate. As $\overline{\lambda}$ increases, the lattice self-organizes into strongly coupled polaronic sites ($\lambda_i \gtrsim \lambda_\star$) whose contribution to $\alpha$ diminishes, and weakly coupled non-polaronic sites ($\lambda_i \lesssim \lambda_\star$) that continue to contribute to $\alpha$ in a renormalized but close to weak-coupling manner. From the balance of these effects emerges the plateau and eventual decline of the $T$-linear slope $\alpha$. We emphasize that our mechanism generates a finite interval of coupling in which $\alpha$ is relatively independent of the coupling.

\vspace{2mm}
\textbf{Matthiessen's rule.}~%
In the entire parameter regime, the scattering rate has the form $\Gamma = \Gamma_0 + \alpha T$. 
Conventionally, Matthiessen's rule would mean that changing $\overline{\lambda}$ simply changes $\alpha$, as it does in weak coupling.
Thus, Matthiessen's rule is dramatically violated in the apparent Planckian scattering regime where increasing $\overline{\lambda}$ leaves $\alpha$ unchanged and instead increases $\Gamma_0$. 
However, this interpretation is only clear because we are privy to the microscopic mechanisms governing the scattering rate. 
In an experimental setting, the exact same macroscopic transport signature of the $T$-linear resistivity translating up as a function of, e.g., irradiation serves as a standard indicator for the \emph{validity} of Matthiessen's rule~\cite{rullier-albenque_influence_2003, ranna_disorder-induced_2025}. 
Our mechanism thus emphasizes that phenomenological conclusions about the validity or violation of Matthiessen's rule should be understood to be predicated on implicit assumptions about the microscopic mechanisms governing the scattering.

\vspace{2mm}
\textbf{Generality of the mechanism.}~%
The essential aspects of our mechanism are not specific to the Holstein model or our particular choice of disordered coupling distribution.
For instance, we expect that introducing a small nonzero mean $\gamma$ will simply drive the system towards the apparent Planckian scattering regime sooner.
Once disorder is introduced into a system, it is natural to consider all parameters as disordered. 
Different sources of disorder will tend to enhance or suppress the metallic state and/or polaron formation, depending on their interplay. 
We expect our mechanism to be valid as long as a metallic state persists (disallowing, e.g., onsite disorder strong enough to induce Anderson localization) and polaron formation is not completely suppressed.
Thus, we conjecture that our mechanism applies to any metallic system of fermions coupled to adiabatic bosons, provided the coupling is spatially disordered, dominates the dynamics, and is sufficient to drive polaron formation and self-trapping.

\vspace{2mm}
\textbf{Future directions.}~%
The most severe limitation of our study is that our concrete verification of the mechanism is in the adiabatic limit $\omega_0 \to 0$. 
It has been argued that introducing finite $\omega_0$ in the clean Holstein model does not qualitatively change the relevant observable properties at $T \gtrsim \omega_0$ \cite{esterlis_breakdown_2018, esterlis_pseudogap_2019}, but it would nevertheless be interesting to simulate the model at finite $\omega_0$, as this would allow access to the regime of inelastic scattering.
We speculate that a suitable generalization of dynamical mean field theory will be sufficient as the results presented here suggest that vertex corrections to the optical conductivity are relatively unimportant and the apparent Planckian scattering is due to local physics. 
It would also be of great interest to explore whether our mechanism, or versions of it, apply in the recent strange metal models based on spatially disordered coupling to critical bosons~\cite{aldape_solvable_2022, patel_universal_2023, patel_strange_2025} or local scatterers~\cite{bashan_tunable_2024,tulipman_solvable_2024}, or in the context of spin polarons~\cite{martinez_spin_1991, bala_spin_1995}.
Finally, we leave a careful study of the implications of our proposed mechanism for the optical conductivity and other observables to future work.

\section*{Conclusion}

We conjectured a general mechanism for the emergence of apparent Planckian scattering $\Gamma_\text{tr} \approx \Gamma_0 + k_B T/\hbar$ driven by local polaron formation. 
We verified this conjecture via Monte Carlo simulations of a version of the paradigmatic Holstein model with disordered electron-phonon coupling, in the adiabatic limit.
In the apparent Planckian regime, we also found a violation of Matthiessen's rule which, however, manifests as a standard experimental signature of the validity of the same rule. 
Ultimately, we have demonstrated how apparent Planckian scattering can robustly emerge in a relatively transparent manner from simple microscopic ingredients.

\section*{Acknowledgements}
We gratefully acknowledge discussions with A.~Pandey, S.~Kivelson, S.~Parameswaran, and H.~R.~Krishnamurthy.
This work was supported by startup funds from the University of Rochester.
The authors thank the Center for Integrated Research Computing (CIRC) at the University of Rochester for providing computational resources and technical support.

\section*{Data availability}
The code used to simulate our model, as well as an example simulation script, is publicly available~\cite{disorderedholstein}.

\bibliography{planck.bib}
\clearpage

\section*{End Matter}

\textbf{A.~The local approximation.}~%
\label{app:LA}
We work with the effective site potentials $q_i = \gamma_i x_i$. In general, the exact distribution of the $q_i$'s is not a product distribution, $P(\vec{q}) \neq \prod_i P_i(q_i)$. Nonetheless, we find that a local approximation $P(\vec{q}) \approx \widetilde{P}(\vec{q}) = \prod_i \widetilde{P}_i(q_i)$ is sufficient to reproduce apparent Planckian scattering. A physically motivated \textit{ansatz} for the local approximation emerges from the atomic limit. In the atomic limit $t_{ij} = 0$, the distribution for $q_i = \gamma_i x_i$ can be computed exactly as
\begin{equation}
\begin{aligned}
    P_i(q_i) = \frac{1}{Z}\biggl[
    &e^{\frac{\beta\nu^2U_i}{2}}e^{-\frac{\beta}{2U_i}(q_i-\nu U_i)^2} \\[-1mm]
    &
    + 2y e^{\frac{\beta(\nu - 1)^2U_i}{2}}e^{-\frac{\beta}{2U_i}[q_i-(\nu-1) U_i]^2} \\
    &
    + y^2 e^{\frac{\beta(\nu - 2)^2U_i}{2}}e^{-\frac{\beta}{2U_i}[q_i-(\nu - 2) U_i]^2} \biggr] ,
    \label{atom}
\end{aligned}
\end{equation}
where $U_i = \gamma^2_i/K$ and $y = e^{\beta \mu}$. The three terms of \eqref{atom} correspond to the atomic states with $n_i = 0, 1, 2$ electrons, respectively. We relax \eqref{atom} to a general mixture of three Gaussians:
\begin{equation}
    \widetilde{P}_i(q_i) 
    = \sum_{a=1}^3 \frac{w_a}{\sigma_a \sqrt{2\pi}} e^{-(q_i-\mu_a)^2/2\sigma_a^2} ,
    \label{LA_distr}
\end{equation}
where $w_1 + w_2 + w_3 = 1$ and $\mu_1 < \mu_2 < \mu_3$,
and fit each local distribution $\widetilde{P}_i(q_i)$ from the MC simulations. Physically, we can interpret this fitting procedure as accounting for the renormalization of various parameters in the atomic limit distribution due to hopping. We end up with a local approximation to the effective site potential distribution $\widetilde{P}(\vec{q}) = \prod_i \widetilde{P}_i(q_i)$ and effective Hamiltonian
\begin{equation}
    H_\text{LA} = -\sum_{ij\sigma}t_{ij}c_{i\sigma}^\dagger c_{j\sigma} + \sum_i q_i n_i ,
\end{equation}
where each $q_i$ is sampled from $\widetilde{P}_i(q_i)$. Simulating the quenched disorder model $H_\text{LA}$ comprises the local approximation. Note that $q_i$ only couples to the electron density $n_i$ and not the shifted density $n_i - \nu$ because the constant shift is taken into account in $\widetilde{P}(\vec{q})$ itself.

Once we have the distributions $\widetilde{P}_i(q_i)$ for the local approximation, we find the coupling $\overline{\lambda}$ at which the first bimodal distribution appears. We detect bimodality by finding two local maxima in the distribution. The coupling that is found represents the onset of local polaron formation. Due to fitting noise, many of the distributions obtained from the local approximation have a component of the Gaussian mixture with a small weight. We must ignore these small weights ($w_i \leq 0.05$) to obtain the correct number of local maxima for a given distribution.

\begin{figure*}[h!]
    \centering
    \includegraphics[width=1.0\linewidth]{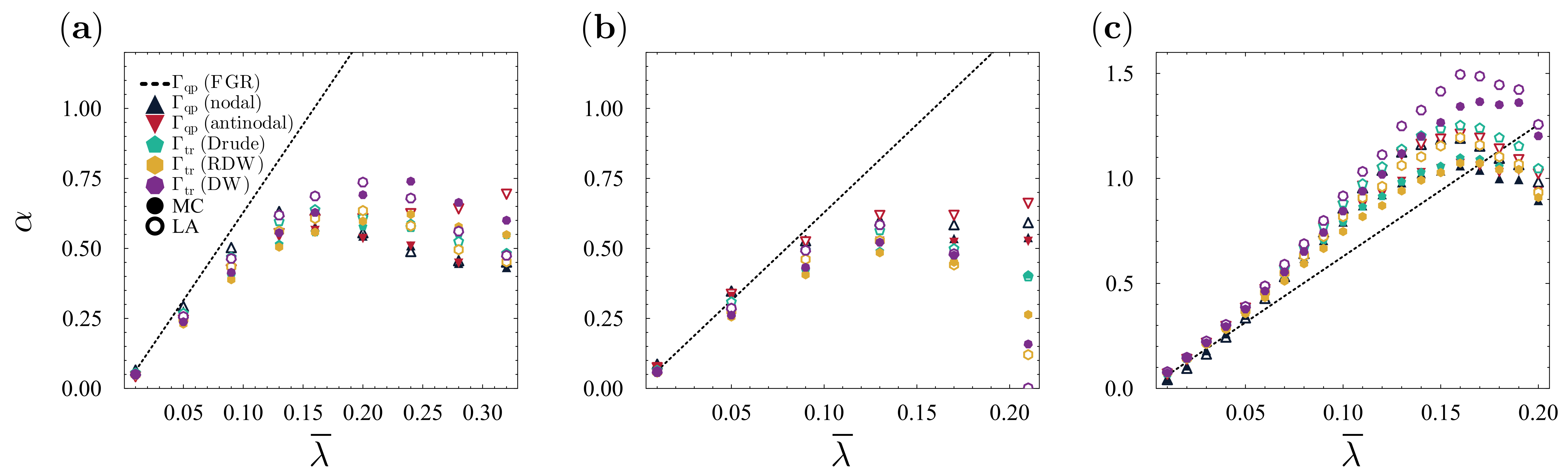}
    \caption{$T$-linear slopes $\alpha$ of scattering rates $\Gamma = \alpha T + \Gamma_0$ (see Fig.~\ref{fig:alpha} and main text for definitions) as a function of $\overline{\lambda}$, for the same model parameters as in Fig.~\ref{fig:alpha} except: \textbf{(a)} at filling $\nu = 1.1$, \textbf{(b)} with band structure set by $t' = 0.2t$, or \textbf{(c)} with local dimensionless couplings distributed according to $\lambda_i \sim \text{Uniform}(0, \lambda_{\text{max}})$.}
    \label{fig:supp_data}
\end{figure*}

\vspace{2mm}
\textbf{B.~Apparent Planckian regime at other model parameters.}~%
\label{app:supp_data}
We verified that the key qualitative features of the results persist if we repeat the calculation with different model parameters, for example with a different filling $\nu = 1.1$, band structure $t' = 0.2t$, or coupling distribution $\lambda_i \sim \text{Uniform}(0, \lambda_\text{max})$. The existence of an apparent Planckian regime for each of these model parameters is shown in Fig.~\ref{fig:supp_data} (compare Fig.~\ref{fig:alpha}b).

\begin{figure*}[h!]
    \centering
    \includegraphics[width=1.0\linewidth]{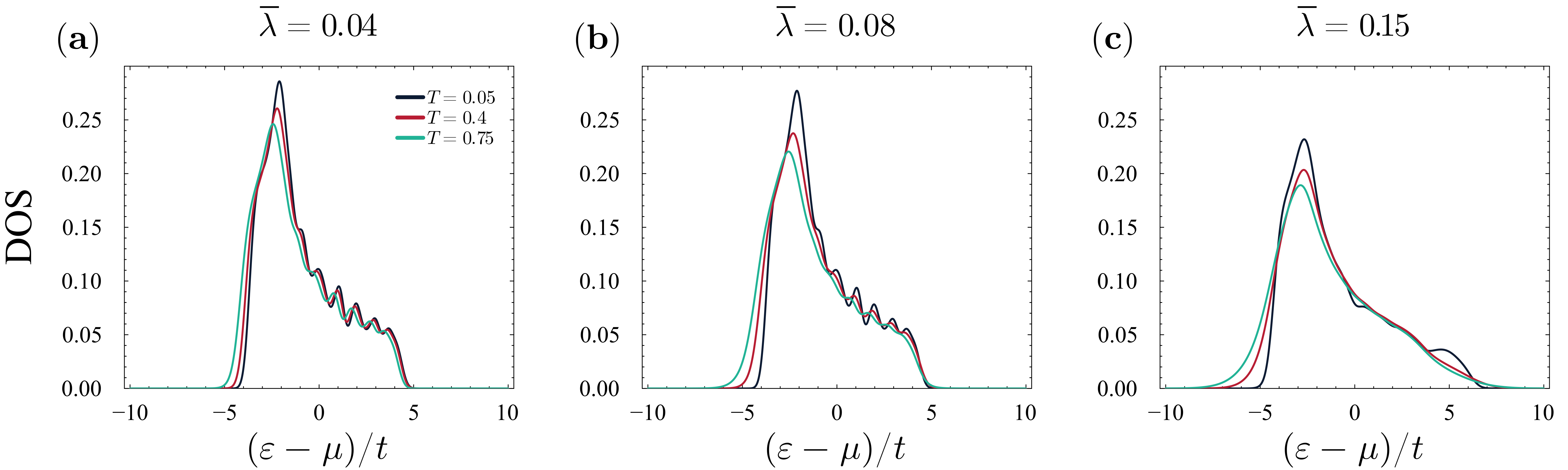}
    \caption{Electronic density of states at representative values of the average coupling $\overline{\lambda}$ and temperature $T$, as a function of energy measured from the chemical potential, $\varepsilon - \mu$, for the same model parameters as in Fig.~\ref{fig:alpha}. The DOS is smoothed with a Gaussian kernel with standard deviation $\sigma = 0.25 t$ to reduce finite size effects.}
    \label{fig:dos+ll}
\end{figure*}

\begin{figure*}[h!]
    \centering
    \includegraphics[width=1.0\linewidth]{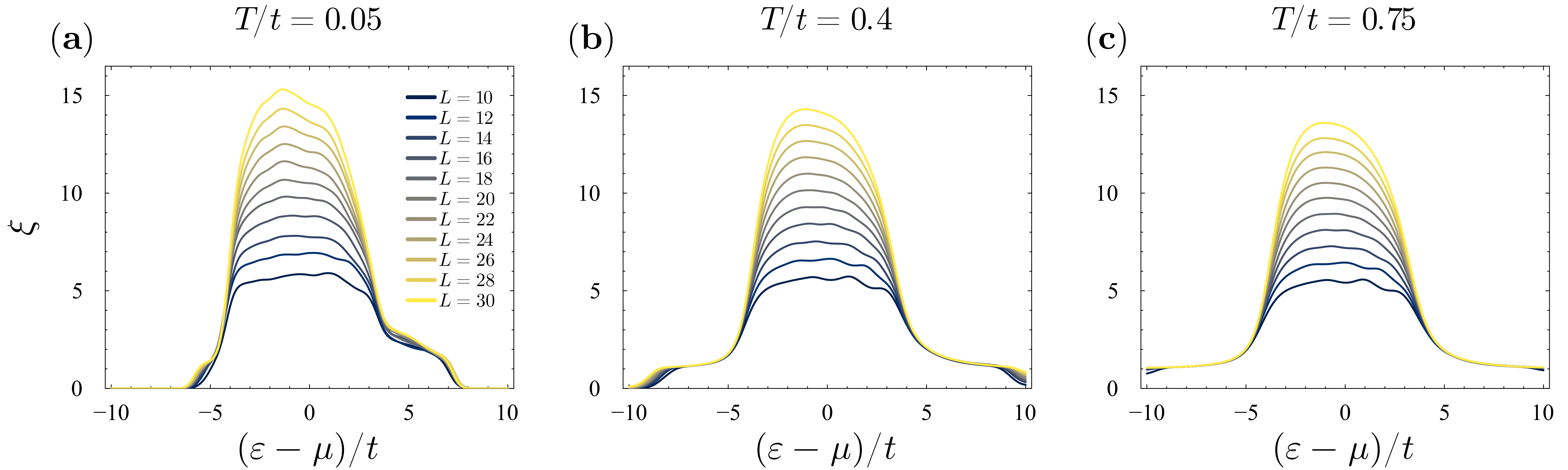}
    \caption{Effective spatial extent (participation length $\xi$) of single particle eigenstates as a function of their energy measured from the chemical potential, $\varepsilon - \mu$, and the linear system size $L$, at average coupling $\overline{\lambda} = 0.15$ for the same model parameters as in Fig.~\ref{fig:alpha}. We smooth $\xi(\varepsilon)$ with a Gaussian kernel with standard deviation $\sigma = 0.25 t$. The $\xi$ of extended states increases with increasing system size, indicating that our results are not affected by Anderson localization.}
    \label{fig:ll_scaling}
\end{figure*}

\vspace{2mm}
\textbf{C.~Density of states and localization length.}~%
\label{app:dos_ll}
As discussed at length in the Supporting Information of Ref.~\cite{murthy_stability_2023}, there are potential subtleties associated with the calculation of dynamical properties in the adiabatic limit $M \to \infty$, having to do with Anderson localization.
To demonstrate that itinerant carriers persist and that our results are not affected by Anderson localization at the system sizes we study, we plot the averaged electronic density of states per spin (DOS) in Fig.~\ref{fig:dos+ll} and the energy-resolved participation length ($\xi$) in Fig.~\ref{fig:ll_scaling}. For a given realization of the couplings $\vec{\gamma}$ and phonon coordinates $\vec{x}$, these are defined as
\begin{align}
    \mathrm{DOS}(\varepsilon;\vec{x},\vec{\gamma}) 
    &= \frac{1}{V} \sum_n \delta(\varepsilon - \varepsilon_n) , \\*
    \xi(\varepsilon;\vec{x},\vec{\gamma})
    &= \sum_n \biggl(\sum_{i} \lvert \langle i \lvert n \rangle \lvert^4 \biggr)^{-1/d} \delta(\varepsilon - \varepsilon_n) ,
\end{align}
where $\lvert n \rangle$ is a spinless single-particle eigenstate of $H_\text{e}(\vec{x}; \vec{\gamma})$ with energy $\varepsilon_n$ and $\lvert i \rangle$ is a position basis state.
We suitably broaden the Dirac distributions and average over phonon configurations (annealed) and couplings (quenched) as in Eq.~\eqref{eq:Oe_average} to obtain $\mathrm{DOS}(\varepsilon)$ and $\xi(\varepsilon)$.

While all single-particle eigenstates are strictly localized in $2$D for static disorder in the thermodynamic limit, we find that, except at energies in the band tails, the average participation length is comparable to the linear system size $L$ and grows (more or less linearly) with it, $\xi \sim L$ (Fig.~\ref{fig:ll_scaling}). This demonstrates that states in the bulk of the band are effectively extended for the range of parameters and system sizes considered (recall that the effective annealed disorder strength increases with $\overline{\lambda}$ and $T$). Moreover, unlike in the clean Holstein model~\cite{murthy_stability_2023}, here no significant pseudogap develops in the electronic DOS at the chemical potential (Fig.~\ref{fig:dos+ll}). Thus, effectively itinerant carriers persist in our simulations and are capable of participating in transport.

\setcounter{secnumdepth}{2}
\appendix
\onecolumngrid 
\clearpage

\section{Monte Carlo algorithm}
\label{app:MALA+ED}

We use the Metropolis-adjusted Langevin algorithm (MALA) to sample phonon configurations. An implementation of this algorithm with exact diagonalization (ED) for the Holstein model is outlined in Ref.~\cite{murthy_stability_2023}. We augment the MALA+ED algorithm with dynamical tuning of the Langevin step size $\tau$ such that the acceptance rate approaches the ideal $0.574$ and the chemical potential $\mu$ such that $\langle N \rangle /V$ approaches $\nu$ during an extended burn-in period. Once the burn-in period ends, the sampling period begins where both $\tau$ and $\mu$ are held fixed and samples are collected. In practice, for each filling $\nu$ we performed a parameter sweep for one realization of the disordered couplings and stored the final values $\{\mu(T, \overline{\lambda}), \tau(T, \overline{\lambda})\}$. We then input this grid of values as the initial guess $\{\mu_0(T, \overline{\lambda}), \tau_0(T, \overline{\lambda})\}$ for all future runs. We find that this procedure effectively eliminates the need for tuning at the cost of running one ``scouting" parameter sweep for each filling. 

Fundamentally, MALA is a Markov chain Monte Carlo (MCMC) algorithm used to sample from a target distribution. To produce correct results, the Markov chain must be sufficiently converged to the distribution of interest and one must collect enough samples to reach tolerable statistical error on the observables of interest. The statistical error can be quantified by the Monte Carlo standard error (MCSE), which is computed from the effective sample size (ESS). The ESS is essentially the number of independent samples that are separated by more than an autocorrelation time, and a larger ESS translates into a smaller statistical error.

The convergence of a Markov chain to the target distribution is more difficult to quantify. The standard practice is to run the same Markov chain from widely dispersed initial conditions and check if all of the chains end up sampling from the same quasi-stationary distribution. Crucially, the quasi-stationary distribution is not necessarily the unique stationary distribution guaranteed in the asymptotic limit $t_\mathrm{MC} \to \infty$ by the fundamental theorem of (ergodic) Markov chains. The Gelman--Rubin statistic $\hat{R}$ quantifies the agreement of the Markov chains started from disparate initial conditions and obeys $\hat{R} \geq 1$ by construction. The condition $\hat{R} \approx 1$ corresponds to all chains sampling from the same quasi-stationary distribution. Thus, $\hat{R} \approx 1$ is a \textit{necessary}, but not sufficient, condition for sampling from the target distribution. The original standard prescribes the threshold $\hat{R} \leq 1.1$ \cite{gelman_inference_1992}, but we adopt a more stringent threshold $\hat{R} \leq 1.01$ as a rule-of-thumb.

In this work, we are primarily concerned with sampling phonon configurations from the target distribution and computing the optical conductivity and electron spectral function with low statistical error. Hence, we report the multivariate $\hat{R}$ for the Markov chain used to sample the full phonon configuration and the MCSE for the optical conductivity and spectral function. We did not find it necessary to perform a rank-normalized split $\hat{R}$ calculation (no chain exhibited a drifting mean) \cite{vehtari_rank-normalization_2021}. Further details on the univariate and multivariate statistics used can be found in Ref.~\cite{vats_revisiting_2021}. We remark that we use the maximum eigenvalue instead of the geometric mean of eigenvalues to compute the multivariate $\hat{R}$ in order to obtain a more conservative statistic.

To compute the statistics, we use the same parameters as in the main text with $T = 0.1t$ and $\overline{\lambda} = 0.15$. We run $10$ chains with a fixed realization of the couplings $\vec{\gamma}$, starting from different initial values of the phonon displacements $x_i$ independently sampled from $\mathcal{N}(0, 64 T/K)$. The chains are run for $5000$ burn steps and $55000$ sample steps. The total number of samples across all chains is thus $N = 550000$. For the Markov chain of phonon configurations, we find $\hat{R}_{\vec{x}} \approx 1.005$. For the Markov chain of longitudinal optical conductivities and spectral functions, we find a negligible MCSE that is not visible when plotted with the data. The maximum MCSE for the optical conductivities, maximized over direction and frequency, is $\text{MCSE}_\sigma \approx 0.001$. The maximum MCSE for the spectral functions, maximized over nodal/antinodal and frequency, is $\text{MCSE}_A \approx 0.0004$.

\clearpage
\section{$T$-linear scattering rates}
\label{app:gamma_T}

\begin{figure}[h!]
    \centering
    \includegraphics[width=1.0\linewidth]{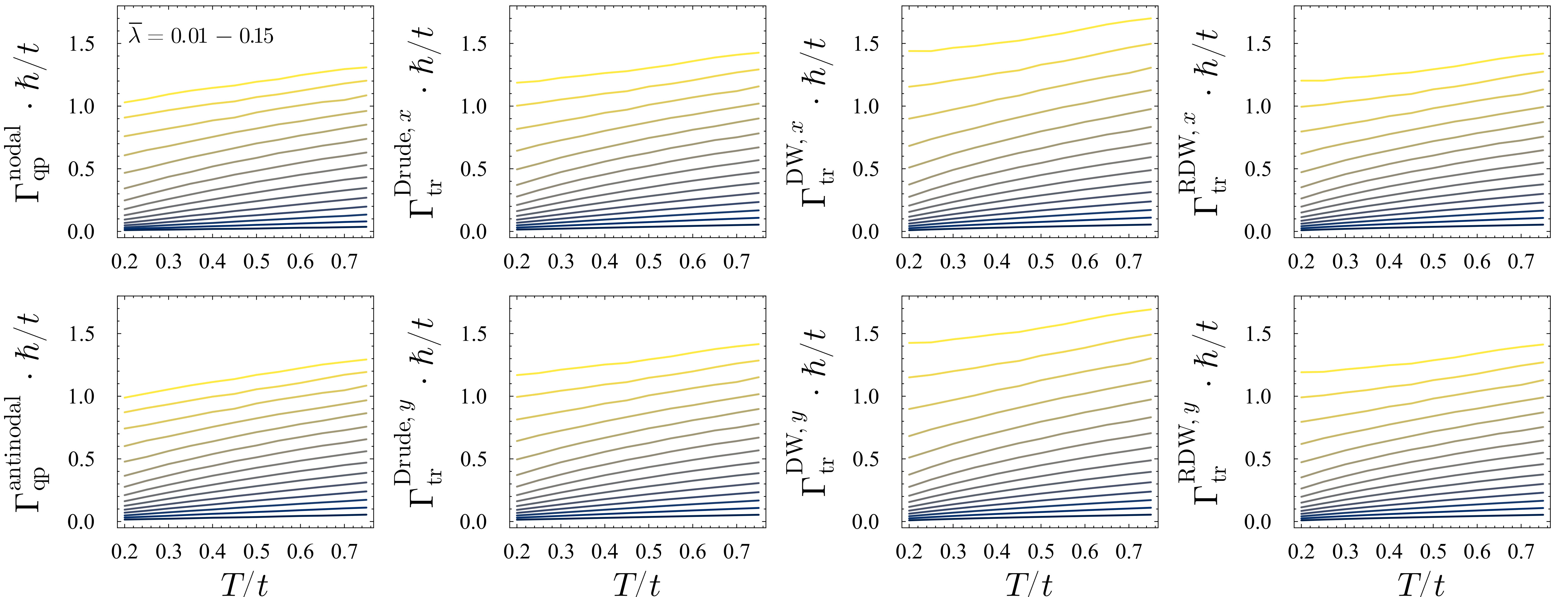}
    \caption{Scattering rates extracted by all methods and for all spatial directions (when relevant) for the 2D square lattice with $L = 20$, $t = 1$, $t' = -0.3t$, $\nu = 1.4$ (same parameters as in the main text). Quasiparticle and transport scattering rates are extracted from the averaged electron spectral function and optical conductivity, respectively. All methods produce similar, approximately $T$-linear scattering rates, which we fit for $T/t \geq 0.4$. The transport scattering rates in the $x$- and $y$-directions also agree with each other, as expected since disorder averaging restores the symmetry of the lattice.}
    \label{fig:gamma_T}
\end{figure}

As discussed in the main text, we use three methods from the existing experimental and theoretical literature on Planckian scattering to extract scattering rates from the quenched and annealed averaged longitudinal optical conductivity $\sigma(\omega)$ and electron spectral function $A(\bm{k}, \omega)$. Figure~\ref{fig:alpha}a in the main text showed that the Drude scattering rate $\Gamma^\text{Drude}_\text{tr}$ is linear as a function of temperature $T$. Figure~\ref{fig:gamma_T} shows that all methods produce similar, approximately $T$-linear scattering rates. Performing a linear-in-$T$ fit on each $\Gamma(T, \overline{\lambda})$ for $T \geq 0.4$ yields the $T$-linear slope $\alpha(\overline{\lambda})$ and $y$-intercept $\Gamma_0(\overline{\lambda})$, which are plotted in Fig.~\ref{fig:alpha}b and~\ref{fig:alpha}c of the main text, respectively.

\clearpage
\section{Electron spectral function, optical conductivity, and scattering rates}
\label{app:spec_opcond}

In the adiabatic limit, the Monte Carlo scheme maps the interacting Hamiltonian $H$ to an ensemble of non-interacting Hamiltonians $H_\text{e}(\vec{x}; \vec{\gamma})$. Thus, for each sample we can use non-interacting expressions to compute the electron spectral function and optical conductivity. The spectral function is given by
\begin{equation}
    A(\bm{k}, \omega; \vec{x}, \vec{\gamma}) 
    = \sum_{n} \lvert \langle n \lvert \bm{k} \rangle \lvert^2 \delta(\omega - \varepsilon_n) ,
\end{equation}
where $\lvert n \rangle$ is a spinless single-particle eigenstate of $H_\text{e}(\vec{x}; \vec{\gamma})$ with energy $\varepsilon_n$, and $\lvert \bm{k} \rangle$ is a spinless single-particle quasimomentum eigenstate. The longitudinal optical conductivity is computed from the Kubo formula
\begin{equation}
\begin{aligned}
    &\sigma_{\alpha\alpha}(\omega; \vec{x}, \vec{\gamma}) 
    = \frac{2\pi}{\omega V}\sum_{nm} \lvert J_{nm}^\alpha \lvert^2 \big[ n_F(\varepsilon_n) - n_F(\varepsilon_{m}) \big] \, \delta(\varepsilon_n - \varepsilon_{m}+\omega) ,
\end{aligned}
\end{equation}
where $\alpha$ indicates spatial direction, $J_{nm} = \langle n \lvert J \lvert m \rangle$ are matrix elements of the current operator in the spinless single-particle eigenstate basis of $H_\text{e}(\vec{x}; \vec{\gamma})$, and $n_F(\varepsilon)$ is the Fermi function. The factor of $2$ is from spin degeneracy. 

\begin{figure}[h!]
    \centering
    \includegraphics[width=0.7\linewidth]{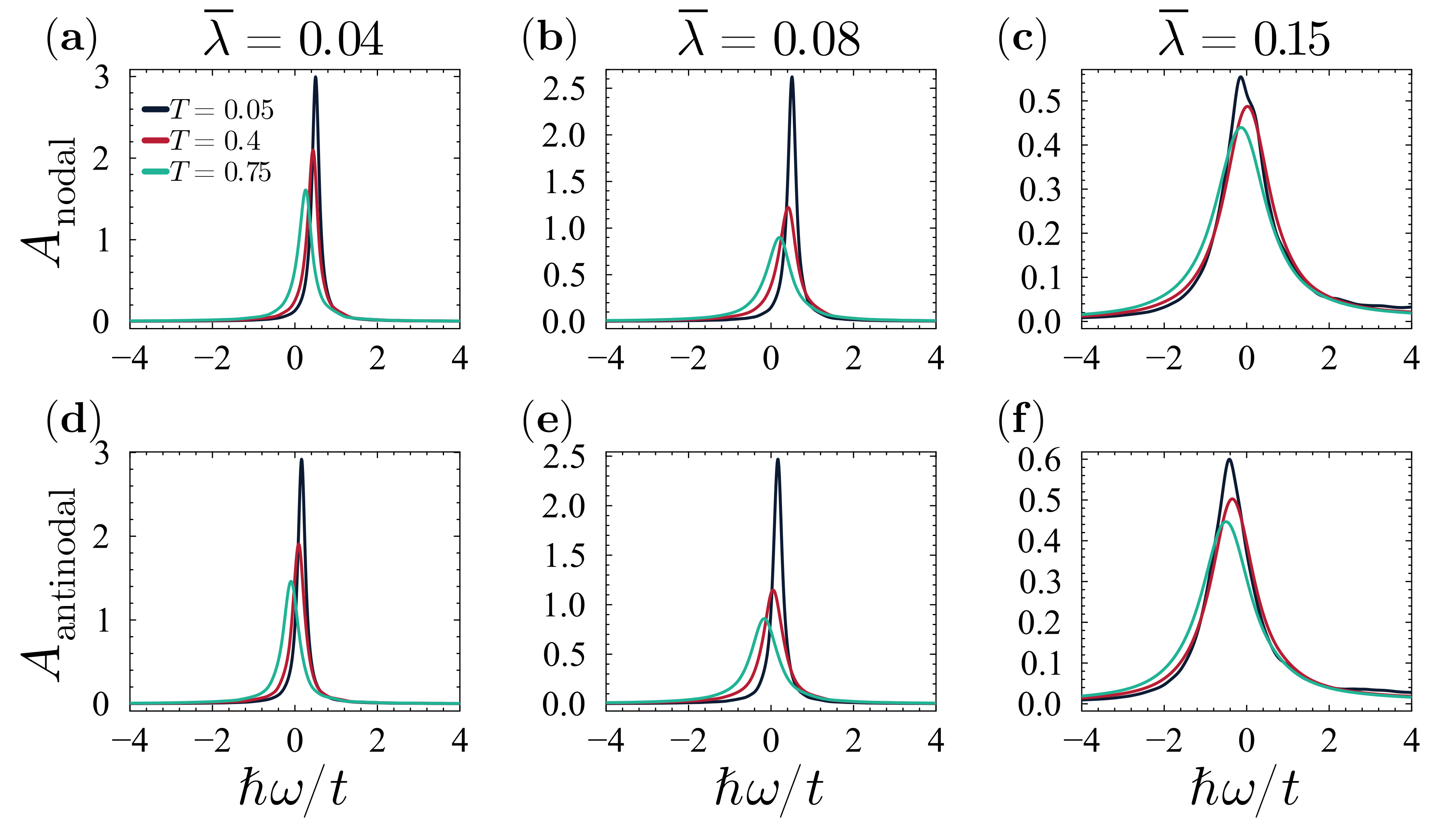}
    \caption{Averaged spectral function evaluated near the nodal ($k_x = k_y$; panels a--c) and antinodal ($k_y = 0$; panels d--f) Fermi surface $\bm{k}$-points at representative values of the average bare coupling $\overline{\lambda}$ and temperature $T$ for the 2D square lattice with $L = 20$, $t = 1$, $t' = -0.3t$, $\nu = 1.4$ (same parameters as in the main text).}
    \label{fig:spectral}
\end{figure}

For both quantities, we broaden the Dirac delta distributions $\delta(x)$ into normalized Lorentzians $\delta_\eta(x) = \eta/\pi(\eta^2 + x^2)$ with broadening $\eta = 2/L$. We have verified that $\eta / \Delta\varepsilon \approx 5$ where $\Delta \varepsilon$ is a typical energy level difference of the single particle spectrum. The non-interacting quantities are then annealed and quenched disorder averaged over phonon displacements $\vec{x}$ and electron-phonon couplings $\vec{\gamma}$, respectively. Representative plots of the spectral function are shown in Fig.~\ref{fig:spectral}, and of the optical conductivity in Fig.~\ref{fig:opcond}.

\subsection{Extraction of scattering rates} 

As discussed in the main text, we use three methods from the existing experimental and theoretical literature on Planckian scattering to extract scattering rates from the averaged electron spectral function $A(\bm{k}, \omega)$ and longitudinal optical conductivity $\sigma(\omega)$. We repeat the discussion below with additional details:

\vspace{2mm}
\textbf{(i) \textit{Nodal/antinodal energy distribution curves}.} 
We compute the spectral function $A(\bm{k}, \omega)$ evaluated near the ``nodal'' ($k_x = k_y$) or ``antinodal'' ($k_y = 0$) Fermi surface $\bm{k}$-points of the bare tight-binding Hamiltonian. Then, we fit the resulting energy distribution curve $A_a(\omega)$ to the usual form of the on-shell spectral function near the (remains of the) Fermi surface:
\begin{equation}
    A_a(\omega) = -\frac{Z_{a}}{\pi}\frac{Z_{a}\text{Im}\Sigma_R(0)}{\omega^2 + [Z_{a}\text{Im}\Sigma_R(0)]^2} ,
\end{equation}
where $Z_{a}$ and $\text{Im}\Sigma_R(0)$ are fitting parameters and $a$ = nodal/antinodal. The quasiparticle scattering rate is then computed as $\Gamma_\text{qp}^a = -2Z_{a}\text{Im}\Sigma_R(0)$. We remark that the choice of $Z_a$ and $\text{Im}\Sigma_R(0)$ as independent fitting parameters yielded better numerical fits than the perhaps more natural choice of $Z_a$ and $\Gamma_\text{qp}^a$ itself. 

\vspace{2mm}
\textbf{(ii) \textit{Drude weight}.}
We compute the Drude weight $D_0$ of the bare tight-binding Hamiltonian by the optical sum rule, $D_0 = \langle\tau\rangle_0/V$, where $\langle \tau \rangle_0$ is the expectation value of the diamagnetic stress tensor in the bare equilibrium state. Then, we extract $\Gamma_\text{tr}^\text{DW} = D_0 / \sigma(0)$. We also compute the \emph{renormalized Drude weight} $D = \langle \tau \rangle / V$, where $\langle \tau \rangle$ is the expectation value in the exact equilibrium state, and extract $\Gamma_\text{tr}^\text{RDW} = D/\sigma(0)$.

\begin{figure}[h!]
    \centering
    \includegraphics[width=0.7\linewidth]{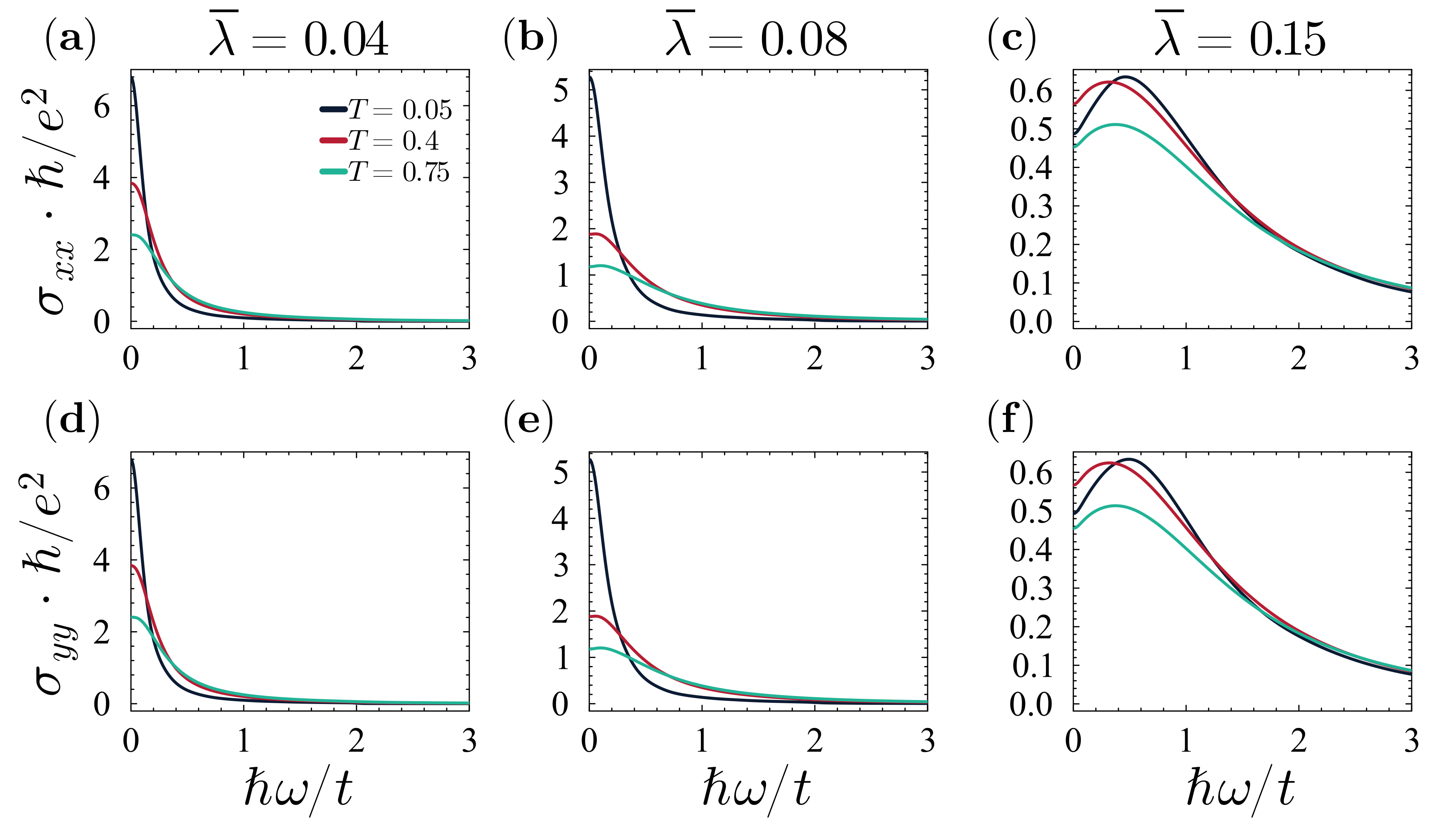}
    \caption{Averaged longitudinal optical conductivity in the $x$-direction (panels a--c) and $y$-direction (panels d--f) at representative values of the average bare coupling $\overline{\lambda}$ and temperature $T$ for the 2D square lattice with $L = 20$, $t = 1$, $t' = -0.3t$, $\nu = 1.4$ (same parameters as in the main text). Note that $\sigma_{xx}(\omega) \approx \sigma_{yy}(\omega)$ as expected since disorder averaging restores the symmetry of the lattice.}
    \label{fig:opcond}
\end{figure}

\vspace{2mm}
\textbf{(iii) \textit{Drude fit}.}
We fit $\sigma(\omega)$ to the Drude form $\sigma_\text{D}(\omega) = \pi D \Gamma_\text{tr} / (\Gamma_\text{tr}^2 + \omega^2)$ and extract $D$ and $\Gamma_\text{tr}$ as fitting parameters. As shown in Fig.~\ref{fig:opcond}, the optical conductivity develops a displaced Drude peak, which has been analyzed in great detail for electron-phonon systems in Ref. \cite{fratini_displaced_2021}. The standard Drude fit is a Lorentzian centered at $\omega = 0$ and thus cannot capture displaced Drude peaks. Nonetheless, the Drude fit yields scattering rate slopes and $y$-intercepts consistent with the same quantities extracted from quasiparticle scattering rates, as shown in the main text. It is ultimately this agreement between all methods that substantiates our finding of an apparent Planckian regime. 


\clearpage

\end{document}